\newcommand{\Fig}[1]{Figure~\ref{#1}}
\newcommand{\alcu}{Al-6.8\%Cu\,}
\documentclass[preprint,12pt,sort&compress]{elsarticle}

\usepackage{geometry}
\usepackage{amssymb}
\usepackage{amsmath}

\usepackage{hyperref} 
\hypersetup{
    colorlinks=true,
    linkcolor=blue,
    citecolor=blue,
    urlcolor=blue,
    filecolor=magenta,
    linkbordercolor={0 0 0}, 
    pdfborder={0 0 0} 
}

\journal{Additive Manufacturing}

\begin{document}

\begin{frontmatter}
\title{Thermal History–Dependent Coalescence Mechanisms and Sintering Dynamics in $\text{Al-}6.8\%\text{Cu}$ Nanopowders}

\author[label1]{Amirhossein Abedini\fnref{equal}}
\author[label2]{Behzad Mehrafrooz\fnref{equal}}
\author[label3]{Iyad Alabd Alhafez\corref{cor1}}\ead{iyad.alabd.alhafez@tu-clausthal.de}
\author[label1]{Arash Kardani\corref{cor1}}\ead{arash.kardani@email.kntu.ac.ir}

\fntext[equal]{These authors contributed equally to this work.}

\cortext[cor1]{Corresponding authors}

\affiliation[label1]{organization={Faculty of Materials Science and Engineering},
            addressline={K. N. Toosi University of Technology}, 
            state={Tehran},
            country={Iran}}

\affiliation[label2]{organization={Center for Computation \& Theory of Soft Materials},
            addressline={Northwestern University}, 
            city={Evanston},
            state={IL},
            country={USA}}

\affiliation[label3]{organization={Institute of Metallurgy, Clausthal University of Technology},
            addressline={Adolph-Roemer Str. 2A}, 
            city={Clausthal-Zellerfeld},
            postcode={38678}, 
            country={Germany}}

\begin{abstract}
Aluminum$-$Copper (Al$-$Cu) alloys are essential materials for weight reduction critical structures in the aerospace and automotive industries, yet achieving their maximum ultrahigh-strength potential remains limited by nanoscale defect control during powder metallurgy processing. We employ large-scale molecular dynamics  simulations on $\text{Al-6.8\%Cu}$ nanoparticles to explore atomic-scale mechanisms governing the full thermal sintering cycle. We demonstrate that while the sintering temperature primarily initiates neck formation, the subsequent cooling rate is the dominant kinetic parameter dictating the final microstructure. Fast cooling rates trap a significantly higher density of stacking faults and can unexpectedly lead to the formation of an amorphous phase at the interparticle interfaces, a feature critically dependent on the rate of thermal dissipation. We confirm a clear shift in the coalescence mechanism from plastic deformation (dislocation slip) at low temperatures ($300 \text{ K}$ and $450 \text{ K}$) to mass transport via atomic diffusion at high temperatures ($600 \text{ K}$). These findings provide essential, atomic-scale guidelines for controlling thermal processing, particularly cooling rates, to design defect-stabilized, high-performance Al$–$Cu components.
\end{abstract}


\end{frontmatter}

\newpage
\section{Introduction}
Aluminum (Al) alloys are foundational of modern lightweight engineering, driving advancements in the aerospace and defense sectors, where high specific strength and thermal stability are critical~\cite{hirsch2008aluminium,rambabu2016aluminium}. Unlike pure aluminum, which is typically limited to non-structural roles, the use of alloying elements provides the  enhanced performance required for advanced structural applications~\cite{georgantzia2021aluminium}. Among these, the 2xxx (Al-Cu) series  stands out as the highest-strength, heat-treatable aluminum family~\cite{das2023state}. The preeminence of the 2xxx series is directly attributable to copper (Cu) being the primary alloying element (typically 2 to 10 wt.\%), which facilitates a crucial mechanism: precipitation hardening~\cite{kenevisi2021review}. During heat treatment, Cu forms the highly effective strengthening phases, notably the metastable Guinier-Preston zones and the stable $\text{Al}_2 \text{Cu}$ precipitates~\cite{kumar2022critical}. While minor elements such as magnesium, manganese, and especially lithium are added to fine-tune specific attributes (\textit{e.g.}, Li reduces density and increases stiffness~\cite{carrick2014influence}), Cu remains the cornerstone for achieving the high tensile strength and superior fatigue resistance characteristic of this alloy group. Optimizing the performance of these alloys is predicated on fine-tuning their internal microstructure, which involves controlling grain size, dislocation density, and the state of these strengthening precipitates~\cite{fan2023review}. Conventional bulk manufacturing methods often impose fundamental limits on achieving ultrafine grain structures or uniform alloying near the phase boundaries. This limitation has driven intense research into powder metallurgy methods~\cite{10.31399/asm.hb.v07.a0006022}, specifically the consolidation of metallic powders through solid-state sintering, as a scalable pathway for producing net-shape components with superior, engineered microstructures~\cite{shrestha2020nanoparticle,babalola2023sintering,hosokawa2012nanoparticle}.

Recent efforts to enhance mechanical properties have shifted toward using nanoparticles (NPs) as sintering precursors.
The extreme surface-to-volume ratio in NPs provides the necessary thermodynamic driving force for rapid densification and grain growth kinetics at lower temperatures, offering a revolutionary approach to materials synthesis [6]. However, this nanoscale regime introduces additional complexities. Atomic-scale mechanisms---including surface atom melting, rapid defect evolution, and neck formation---are highly sensitive to processing conditions and are inaccessible to conventional \textit{in situ} experimental characterization. Consequently, Molecular Dynamics (MD) simulations have emerged as an indispensable tool for mapping these processes~\cite{zhu1996sintering,ding2009molecular,yi2023modeling,kardani2018md,wang2022study,meng2022computational,sun2025coalescence}, successfully revealing how parameters like sintering temperature~\cite{malti2021insight,wang2024effect,abedini2022mechanical,jamshideasli2024molecular,meng2022computational,rahbar2023sintering,jiang2020monitoring}, applied pressure~\cite{hu2020thermal,lyu2025low,kim2024atomistic}, surface roughness~\cite{liu2025effect}, and particle size~\cite{jamshideasli2024molecular,nandy2019sintering,abedini2023probing,song2022atomic,rahbar2023sintering} dictate initial coalescence and mechanical behavior in metallic systems. 
Despite these advances, a critical and industrially relevant gap remains: the influence of the full thermal cycle. Manufacturing processes do not stop at the isothermal hold; the subsequent cooling stage is equally deterministic of the final microstructure~\cite{castro2012sintering}. In Al-Cu systems, the cooling rate dictates the degree of dislocation recovery, the formation of vacancies, and, crucially, the kinetic window for the nucleation and growth of strengthening precipitates like $\text{Al}_2 \text{Cu}$, which are known to strongly influence the final yield strength and ductility. The atomic-level coupling between the post-sintering cooling rate and the stabilization of the nanoscale defect structure and bonding quality in high-performance Al-Cu alloys has not been systematically investigated. Addressing this gap is essential for translating MD insights into practical, high-quality component manufacturing.

In this study, we use large-scale, high-fidelity MD simulations to systematically elucidate the complete thermal sintering pathway for Al-6.8\%Cu alloy NPs. The concentration of $6.8 \text{ wt.\%}$ $\text{Cu}$ is near the maximum solid solubility of Cu in Al at the eutectic temperature; this specific composition is chosen to ensure the simulation models the highest-performance binary variant, capable of producing the greatest potential volume fraction of strengthening $\text{Al}_2 \text{Cu}$ precipitates. Uniquely, we characterize the combined, two-stage influence of the sintering temperature (isothermal hold) and, for the first time in this system, the cooling rate on the microstructural evolution of the final sintered compact. We quantify key structural metrics including density, void fraction, and surface area, and employ two analysis tools; Common Neighbor Analysis (CNA)~\cite{honeycutt1987molecular} and the Dislocation Extraction Algorithm (DXA)~\cite{stukowski2010extracting}, to precisely track the evolution of the crystalline phases and defect populations. Our findings demonstrate that while sintering temperature initiates neck formation, the cooling rate is the dominant kinetic parameter during structural relaxation, controlling dislocation recovery and the resulting atomic arrangement at sintered neck interfaces. This research provides fundamental atomic-scale insight and establishes a comprehensive, predictive set of kinetic guidelines for tailoring the microstructure and optimizing the performance of 2xxx series Al-Cu alloys fabricated via advanced powder metallurgy techniques.

\section{Methodology}
\label{sec:simdetails}
\begin{figure}[!t]
    \centering
    \includegraphics[width=\textwidth]{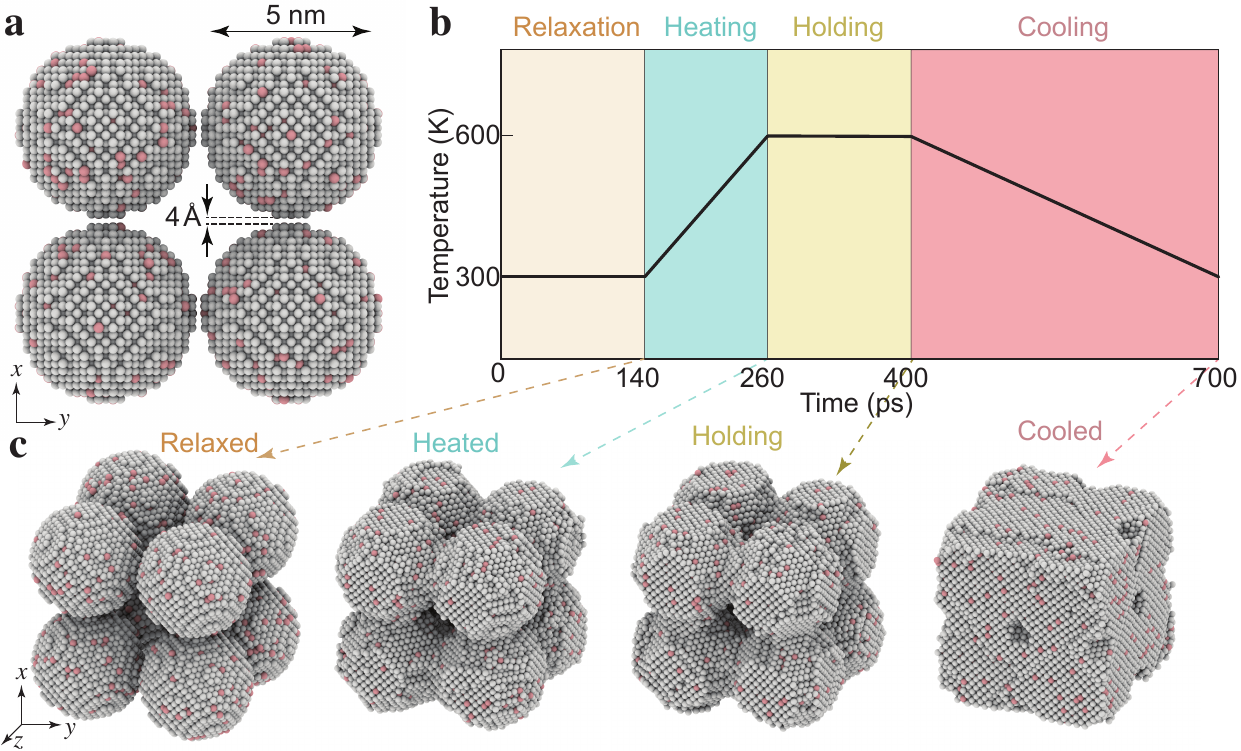}
    \caption{\textbf{All-atom model of the \alcu\ nanoparticles.} 
    \textbf{a,} Top view of the initial configuration consisting of eight 5-nm-diameter \alcu\ spheres separated by 4~\AA. Aluminum atoms are shown in gray and copper atoms in pink. 
    \textbf{b,} The complete thermal profile used for the sintering simulation. The cycle includes initial structural relaxation at 300~K, heating to the sintering temperature of 600~K, a critical isothermal hold period, and the final controlled cooling stage. Snapshots beneath the plot illustrate the system's evolving configuration at the end of each major thermal stage.}
    \label{fig1}
\end{figure}

The initial \alcu NP assembly was constructed using the Atomsk package \cite{hirel2015atomsk}. The final model comprised eight spherical \alcu NPs, each with a diameter of $5 \text{nm}$ \cite{wang2024effect,abedini2022mechanical,abedini2023probing}, shown in \Fig{fig1}a,c. The $\text{Al-Cu}$ alloy system was modeled as a face-centered cubic (FCC) structure with a lattice constant of $4.049 \text{ \AA}$. The $\text{Cu}$ atoms were incorporated into the $\text{Al}$ matrix via random substitution of $\text{Al}$ atoms to achieve the target weight fraction of $6.8\%$. This resulted in a system containing $31,072$ total atoms, specifically $28,958$ $\text{Al}$ atoms and $2,114$ $\text{Cu}$ atoms. To prevent initial atomic overlap, a uniform $4 \text{ \AA}$ gap~\cite{ji2025structural} was set between adjacent NPs (\Fig{fig1}a). The entire system was simulated with periodic boundary conditions (PBCs) applied in all three directions.

All MD simulations were performed using the Large-scale Atomic/Molecular Massively Parallel Simulator (LAMMPS) software package \cite{thompson2022lammps}. Atomic interactions between $\text{Al}$ and $\text{Cu}$ atoms were accurately described using the Embedded Atom Method (EAM) potential developed by Cai \textit{et al}.~\cite{cai1996simple} which accurately reproduces the cohesive, elastic, and thermodynamic properties of FCC Al–Cu alloys. This potential has been extensively validated in prior MD studies of Al–Cu systems~\cite{abedini2022mechanical,abedini2023probing}, ensuring reliable representation of both solid-state and partially molten configurations during the simulated thermal cycle".
The time step for all integrations was set to $1 \text{ fs}$. 
Temperature control throughout the sintering and cooling stages was maintained using the Nose-Hoover thermostat~\cite{evans1985nose}. Simulations were conducted primarily in the isothermal–isobaric ensemble ($\text{NPT}$) to control temperature effectively.

Each simulation followed a consistent three-stage thermal cycle. As plotted in \Fig{fig1}b, Initially, the configurations underwent a relaxation period of $140 \text{ ps}$ under the canonical ensemble to ensure stability and minimize potential energy prior to sintering. The sintering phase was initiated by heating the relaxed NPs at a rate of $2.5 \text{ K/ps}$ \cite{liu2023coalescence,liu2020sintering,singh2021nano,li2023cu} to target temperatures of $300 \text{ K}$, $450 \text{ K}$, and $600 \text{ K}$, chosen to investigate the transition between plastic deformation and diffusion-dominated coalescence. This heating was followed by an isothermal hold under $\text{NPT}$. Finally, the assemblies were subjected to a controlled cooling stage, returning to $300 \text{ K}$ at three distinct rates: $0.1 \text{ K/ps}$, $1 \text{ K/ps}$, and $10 \text{ K/ps}$ \cite{wang2024effect,ren2020amorphous,luo2021crystallization,zhou2025amorphization}, the variation being critical for assessing the kinetic trapping of defects and the evolution of the final microstructure. Table~\ref{tab:sintering_params} summarizes the simulation parameters for used for spark plasma sintering, heating and cooling rates employed in the simulations.

Microstructural analysis was primarily conducted using the OVITO software package~\cite{stukowski2009visualization}. Visualization of the atomic configurations was performed using the VMD software package~\cite{humphrey1996vmd}. Two key atomistic analysis tools were employed: the CNA~\cite{honeycutt1987molecular} method, used to distinguish the crystalline phases from the amorphous phase and to identify crystalline defects such as $\text{Stacking Faults (SFs)}$; and the DXA~\cite{stukowski2010extracting}, utilized to characterize the presence, density, and Burgers vectors of dislocations throughout the sintering process. Additional metrics, including surface area, void fraction, and density, were calculated to quantify the extent of sintering progression and densification.
Surface mesh analysis was employed to quantify changes in surface area during sintering across different temperatures. The void fraction was defined as the ratio of the unoccupied volume of the simulation box to its total volume.

\section{Results and Discussion}

\subsection{Thermal Stability and Melting Behavior of $\text{Al-6.8\%Cu}$ Nanoparticles}

\begin{figure}[!t]
    \centering
    \includegraphics[width=\textwidth]{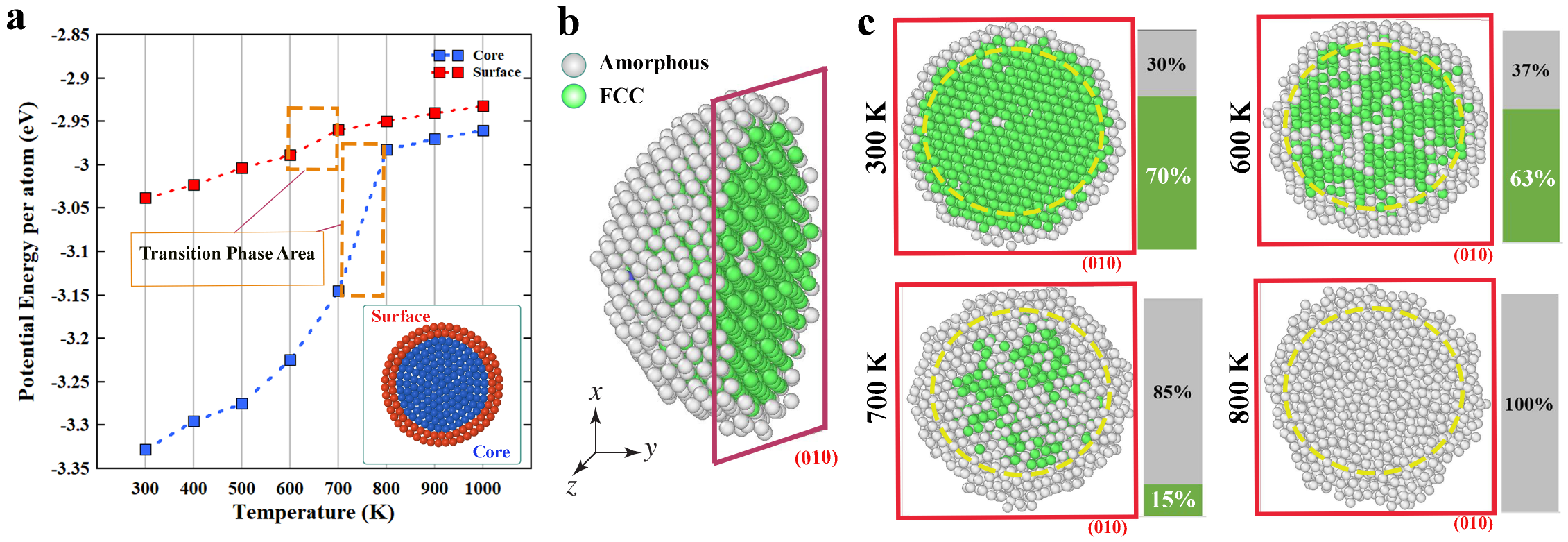}
    \caption{\textbf{Melting behavior of surface and core regions in a \alcu NP.}
    \textbf{a,} Temperature dependence of the potential energy per atom. 
    Surface atoms (red) are defined as those within 2.5~\AA\ of the outer surface, while the core (blue) consists of the remaining atoms. 
    \textbf{b,} Cross-sectional view of a \alcu NP with a radius of 2.5~nm, sliced along the (010) plane. 
    \textbf{c,} Final atomic configurations at different temperatures (300~K, 600~K, 700~K, and 800~K), showing FCC atoms in green and amorphous atoms in gray.
    The side bars indicate the relative fraction of FCC and amorphous atoms at each temperature, highlighting the progressive loss of crystalline order during heating.}
    \label{fig2}
\end{figure}

Prior to analyzing the coalescence and microstructural evolution during the active sintering phase, it is essential to establish the thermal stability limits of \alcu  NPs. The material must be characterized across the temperature range of the sintering procedure to understand the temperatures at which structural degradation begins, a crucial precursor to bonding. To precisely determine the melting points of a $5 \text{ nm}$ \alcu NP, we started by simulating a single \alcu NP and tracked the potential energy per atom ($\textit{E}_{\text{pot}}$) as a function of temperature.  We divided the atoms into distinct surface and core regions to reveal the progression of the phase transition. Atoms within $2.5 \text{ \AA}$ of the NP surface were defined as surface atoms, with all remaining atoms considered as core~\cite{wang2024effect}. 
\Fig{fig2}a plots $\textit{E}_{\text{pot}}$ for the defined core and surface regions, revealing two distinct regions of phase transition.
The surface atoms exhibit a sharp increase in potential energy between $600 \text{ K}$ and $700 \text{ K}$, indicating the initiation of melting with a defined surface melting point of approximately $650 \text{ K}$. Conversely, the core atoms show a delayed phase change, with a clear jump in $\textit{E}_{\text{pot}}$ occurring between $700 \text{ K}$ and $800 \text{ K}$, establishing the core melting point ($T_{\rm{m}}$) around $750 \text{ K}$. In agreement with previous literature, our simulations confirm the phenomenon of melting point depression at the nanoscale, where the large specific surface area drives a continuous decrease in the melting temperature compared to the bulk material~\cite{jiang2006size}.

This shell-to-core melting behavior is further structurally confirmed by monitoring the loss of the FCC crystal structure within the NP as temperature increases,~\Fig{fig2}b. At the initial temperature ($300 \text{ K}$), approximately $70\%$ of atoms possess a stable $\text{FCC}$ structure, distributed primarily in the core, while the remaining $30\%$ form an amorphous phase confined to the surface shell (\Fig{fig2}c). As the temperature is increased toward the surface melting point, the $\text{FCC}$ atoms rapidly transition into the liquid-like amorphous phase, driven by the increased potential energy. The fraction of amorphous atoms increases to $37\%$ at $600 \text{ K}$ and jumps sharply to $85\%$ at $700 \text{ K}$ (\Fig{fig2}c), demonstrating that nearly all surface atoms have converted by $700 \text{ K}$. The presence of this amorphous, liquid-like state is verified by the Radial Distribution Function (RDF) analysis (\Fig{SI_rdf}), which confirms short-range atomic order ($r \approx 2.8\text{--}3.0$~\AA) combined with long-range disorder (broadening of the second, third, and fourth peaks) \cite{huang2024evolution}. This progressive liquefaction, starting at the surface ($650 \text{ K}$) and propagating inward to the core ($750 \text{ K}$), confirms the shell-to-core melting mechanism and dictates the thermal range available for solid-state sintering.

\subsection{Microstructural Evolution During Sintering}
\begin{figure}[!t]
    \centering
    \includegraphics[width=0.9\textwidth]{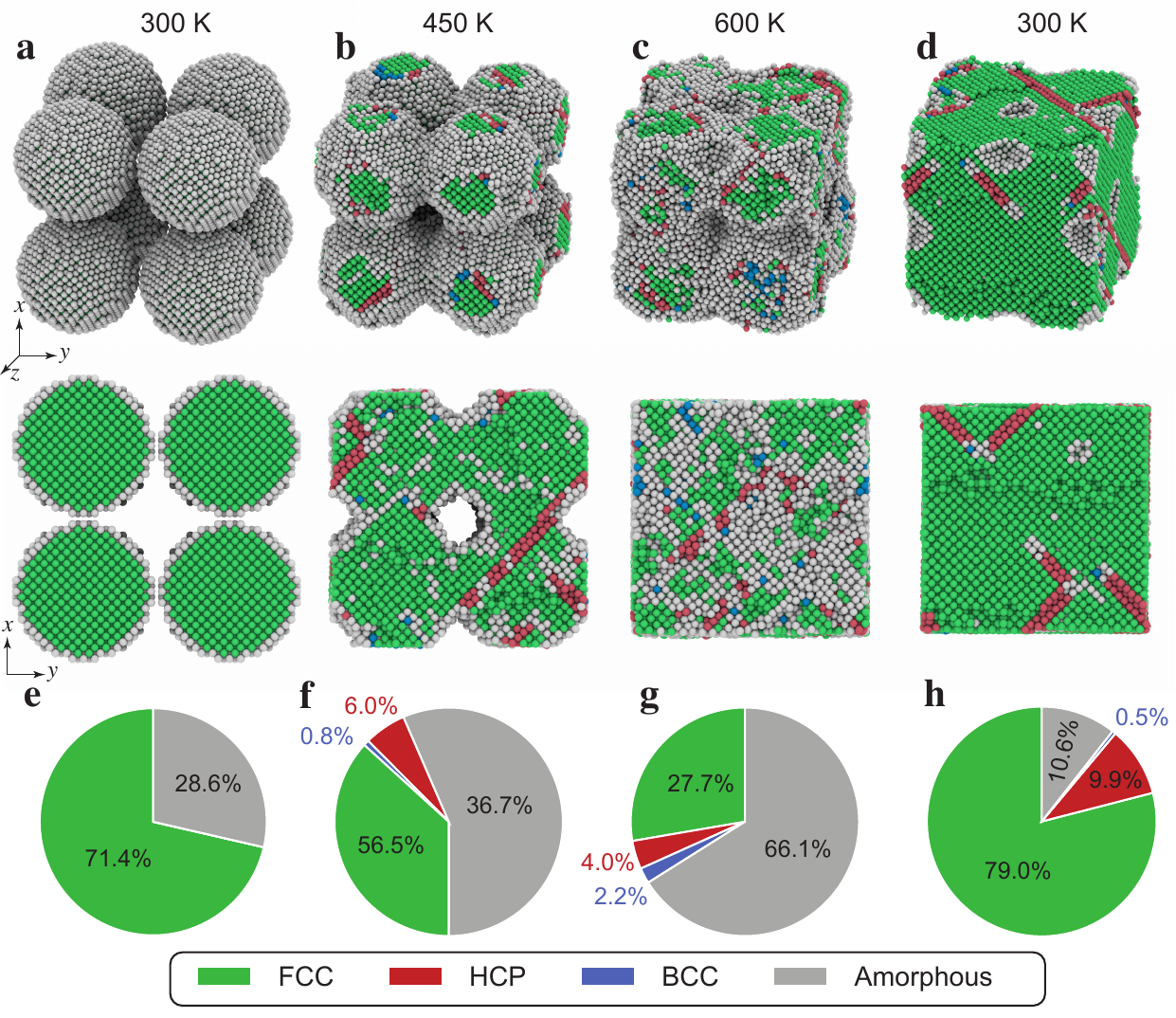}
    \caption{\textbf{Microstructural evolution of \alcu NPs during sintering.} 
    \textbf{a–d,} Atomic configurations of \alcu NPs during annealing process up to 600 K shown in three-dimensional (top row) and (010)-plane cross-sectional (bottom row) views at: (a) the beginning of the relaxation stage, (b) mid-heating stage, (c) mid-holding stage, and (d) the end of the cooling stage. 
    The (010)-plane section is taken through the center of the NPs. 
    Hereafter, atoms and plots are color-coded according to their local crystal structure: FCC (green), HCP (red), BCC (blue), and amorphous (gray).
    \textbf{e–f,} Corresponding phase composition represented as pie charts showing the fraction of each structural type for the snapshots in (a–d).}
    \label{fig3}
\end{figure}
The ultimate mechanical and physical properties of sintered materials are intrinsically linked to their microstructural evolution during thermal processing. To elucidate these dynamic changes at the atomic level, we employed CNA to track the crystallographic phase transformations in \alcu NPs throughout a simulated sintering cycle. Initially, the NPs exhibit a core-shell like structure, with a crystalline FCC core comprising approximately 71.4\% of the atoms, encapsulated by a 28.6\% amorphous surface layer (\ref{fig3}a,e). The beginning of heating initiates two concurrent phenomena: the physical aggregation of NPs and a profound structural transformation. As thermal energy increases atomic mobility, the primary FCC phase begins to destabilize, its fraction decreasing from 71.4\% to 56.5\%. This destabilization primarily fuels the growth of the amorphous phase, which expands to 36.7\%. Concurrently, localized stresses induced by particle contact and plastic deformation trigger the nucleation of Hexagonal Close-Packed (HCP) atoms, which manifest as stacking faults and constitute 6\% of the structure. A minor fraction of Body-Centered Cubic (BCC) atoms (0.8\%) also emerges (\Fig{fig3}b,f). Physically, this stage is characterized by the initial contact between NPs and the formation and subsequent growth of a sintering neck (\Fig{SI_necking}).
Upon reaching 600~K, the crystallographic arrangement is substantially altered. During the holding stage, sufficient thermal energy and time allow the system to approach a quasi-equilibrium state dominated by disorder. The amorphous phase fraction surges to a maximum of 66.1\%, while the primary FCC phase diminishes to just 27.7\% (\Fig{fig3}c,g). This extensive amorphization is driven by the entropic favorability of disordered states at high temperatures. The HCP phase also reduces to 4\%, suggesting that it is a transient structure that readily transforms into the amorphous phase under these conditions. In contrast, the BCC fraction increases slightly to 2.2\%, indicating its comparatively greater thermal stability. At this point, the NPs have coalesced into a nearly fully aggregated structure with a well-developed sintering neck, \Fig{fig3}c.
The final stage of the process, cooling, acts as a structural quench, driving the system back toward a low-energy, ordered configuration. The reduction in thermal energy removes the driving force for amorphization, triggering a rapid recrystallization event. The amorphous phase fraction plummets from 66.1\% to just 10.6\%. This is mirrored by a substantial increase in the FCC and HCP phases, which reach final concentrations of 79.0\% and 9.9\%, respectively (\Fig{fig3}d,h). Notably, the final structure is more crystalline than the initial state, indicating that the thermal cycle not only consolidates the NPs but also serves as an annealing process that enhances overall crystallinity. The thermally stable BCC phase, no longer favorable at lower temperatures, largely transforms into FCC and HCP structures, its fraction drops to a negligible 0.5\%.

\begin{figure}[!t]
    \centering
    \includegraphics[width=0.8\textwidth]{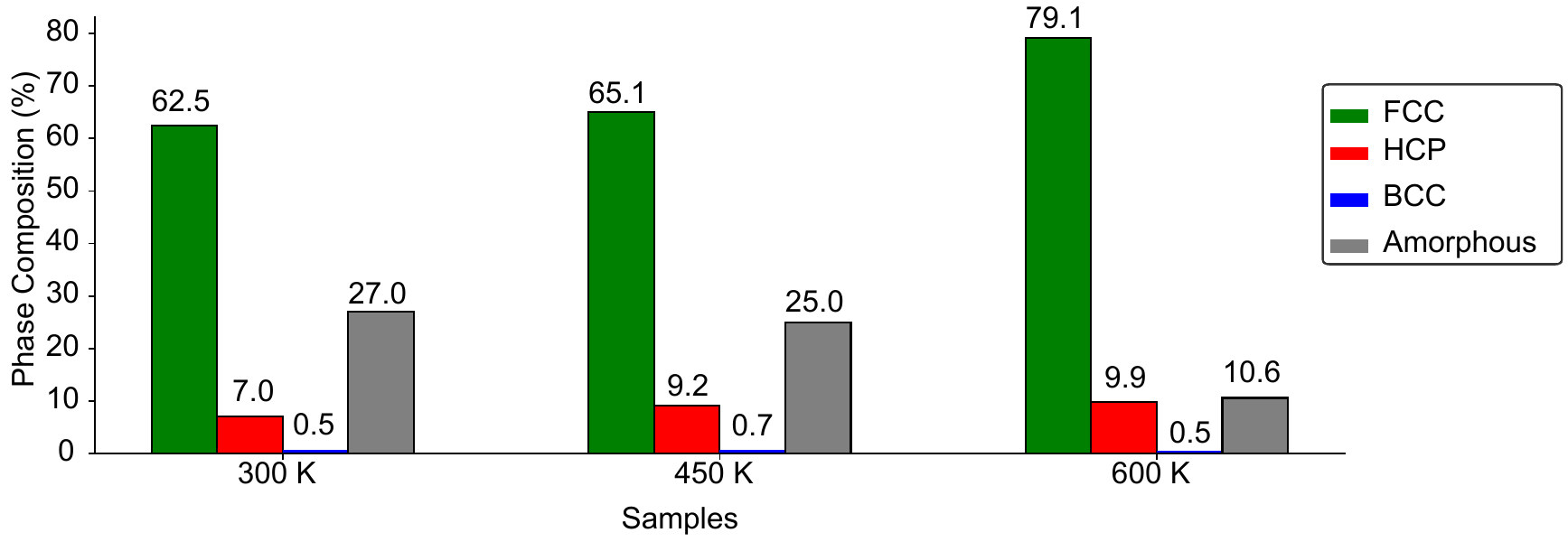}
    \caption{\textbf{Temperature-dependent microstructural evolution of sintered \alcu NPs.} 
    Relative fractions of FCC, HCP, BCC, and amorphous phases extracted at the end of the sintering process for simulations performed at 300, 450, and 600~K, corresponding to approximately 0.4, 0.6, and 0.8 of the average melting point of the core atoms.}
    \label{fig4}
\end{figure}

To further probe the role of the observed amorphization-recrystallization pathway, we extended our simulations to systematically investigate how the final microstructural landscape is governed by the thermal budget. We performed comparative analyses at peak sintering temperatures of 300~K, 450~K, and 600~K, corresponding to homologous temperatures of approximately $0.4\,T_{\rm{m}}$, $0.6\,T_{\rm{m}}$, and $0.8\,T_{\rm{m}}$, respectively. 
The temperature of 300 K was chosen specifically to examine the process under ambient conditions, as NPs are capable of sintering at room temperature~\cite{magdassi2010triggering,yang2023evolving}.
The results, summarized in \Fig{fig4}, reveal that the ultimate crystallographic texture is highly sensitive to the maximum processing temperature.
A direct, positive correlation is observed between the sintering temperature and the final degree of crystallinity. As the temperature increases from 300~K to 600~K, the volume fractions of both the primary FCC phase and the associated HCP stacking faults increase monotonically. This enhancement of ordered structures signifies that greater thermal energy provides a more effective kinetic pathway for annealing out initial surface defects and promoting crystallographic perfection during consolidation.
Conversely, the fraction of the residual amorphous phase is inversely proportional to the peak temperature. The significant reduction in amorphous content at 600~K is driven by a dual mechanism. First, higher temperatures promote more complete particle coalescence, which inherently minimizes the total surface area—the primary location of the initial disordered atoms. Second, and more critically, the extensive transient amorphization at 600 K creates a liquid-like precursor state. During cooling, the enhanced atomic mobility from this state provides superior kinetic conditions for atoms to rearrange into the thermodynamically favored crystalline lattice, more effectively preventing the disorder from being quenched into the final structure. The minor BCC phase remained nearly constant throughout the sintering process, with fractions of 0.5–0.7\%, suggesting that these values are within the expected thermal or vibrational fluctuations and do not indicate significant structural changes.

\subsection{Thermal Control of Nanoparticles Consolidation and Densification}
\begin{figure}[!t]
    \centering
    \includegraphics[width=0.9\textwidth]{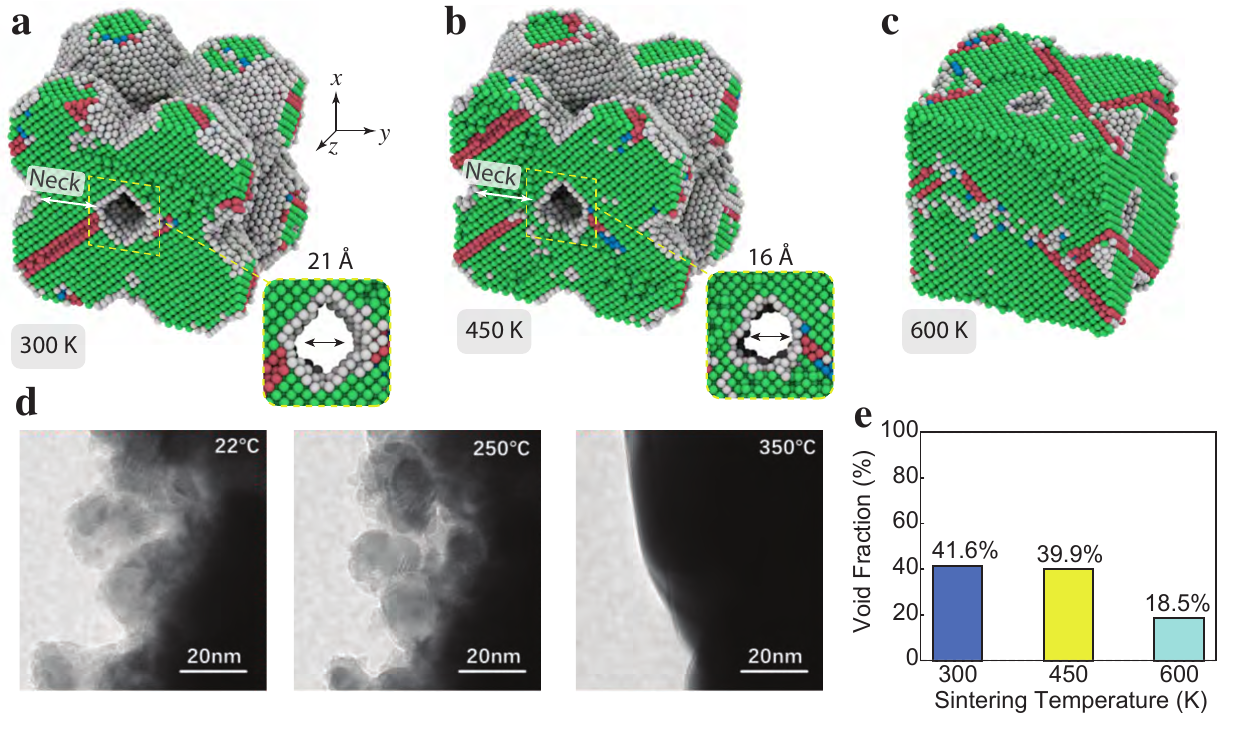}
    \caption{\textbf{Final microstructures and void fraction of \alcu NPs sintered at various temperatures.} 
    \textbf{a,} Cut-away view of the NP microstructure at the center of the sphere at 300 K, 
    \textbf{b,} 450 K, and 
    \textbf{c,} 600 K. The inset shows a cross-sectional view of the void region between eight NPs, with the sintering neck highlighted by a white arrow. 
    \textbf{d,} In-situ TEM images of the sintering neck at 22 °C (295 K), 250 °C (523 K), and 350 °C (623 K), reproduced with permission from \cite{liu2023coalescence} (Elsevier). 
    \textbf{e,} Void fraction of the NPs, calculated as the total NP volume divided by the simulation box volume.}
    \label{fig5}
\end{figure}

\begin{figure}[!t]
    \centering
    \includegraphics[width=\textwidth]{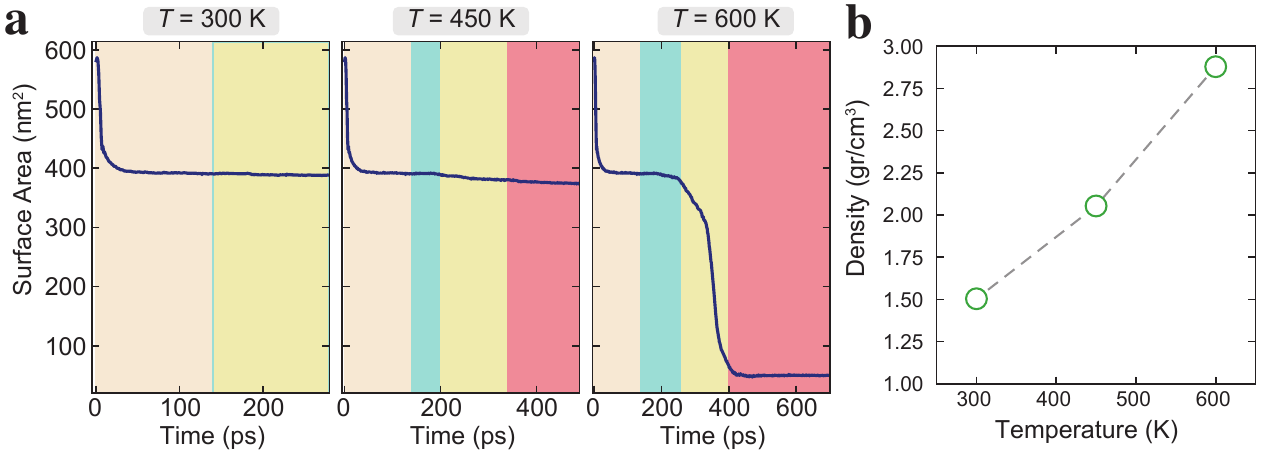}
    \caption{\textbf{Evolution of surface area and density of \alcu NPs during sintering.} 
    \textbf{a,} Evolution of surface area and density of Al–6.8\%Cu NPs during sintering at 300, 450, and 600 K. 
    \textbf{b,} Variation of NP density as a function of sintering temperature, highlighting the progressive densification with increasing thermal energy.}
    \label{fig6}
\end{figure}

To forge a continuous, dense solid from discrete NPs, the sintering process must overcome significant kinetic and thermodynamic barriers. So far, our simulations reveal that the sintering temperature is the critical parameter governing this structural transformation, primarily by controlling a transient, liquid-like phase that dictates mass transport.
The morphological evolution provides a clear visual narrative of this process. At 300~K, NPs contact is minimal, characterized by nascent neck widths of approximately 30~\AA  (\Fig{fig5}a). Increasing the temperature to 450~K promotes further neck growth up to 36~\AA, yet the particles retain much of their individual character (\Fig{fig5}b). A dramatic transition occurs at 600~K , where the NPs coalesce into a nearly fully integrated structure (\Fig{fig5}c). Crucially, the sintering necks at the particle-particle interfaces are dominated by the amorphous phase, confirming its role as the primary conduit for atomic diffusion and bonding. This computational result is in excellent qualitative agreement with experimental high-resolution TEM observations~\cite{liu2023coalescence} of sintered Cu NPs, which similarly demonstrate a transition from partial necking to full integration with increasing temperature (\Fig{fig5}d). This visual consolidation is further quantified by a calculation of NPs densification, \Fig{fig5}e. The void fraction, a key determinant of material properties, exhibits a striking dependence on temperature. While modest densification occurs between 300~K (41.6\% void) and 450 K (39.9\% void), a profound collapse of the void space is observed at 600~K, where the fraction plummets to 18.5\%. This represents a more than two-fold increase in packing efficiency.
This densification occurs as the system seeks to reduce excess energy, which is physically reflected in the substantial decrease in total surface area.
As shown in \Fig{fig6}a, the system's surface area decreases only slightly at lower temperatures. However, at 600 K, the surface area undergoes a rapid, order-of-magnitude collapse from $\sim$390 nm$^2$ to just $\sim$40 nm$^2$. This is the macroscopic signature of the enhanced atomic diffusion enabled by the transient, high-temperature amorphous phase, which directly leads to a higher final density (\Fig{fig6}b).

In conclusion, the sintering temperature dictates the final physical architecture of the nanomaterial by activating a specific mechanism. A critical thermal threshold must be crossed to induce a transient, liquid-like state that facilitates rapid mass transport. This mechanism is responsible for accelerating void elimination~\cite{kim2024atomistic}, minimizing surface energy, and ultimately forming a dense, consolidated structure. 

\subsection{A Mechanistic Crossover from Plasticity to Diffusion Governs Nanoparticle Sintering}
While several pathways including surface diffusion, grain-boundary diffusion, lattice diffusion, plastic flow through dislocation slip, and vapor transport~\cite{herring1950effect,grammatikopoulos2014coalescence,shi2021towards} can theoretically govern sintering, the process at the nanoscale is dominated by a tug-of-war between two primary mechanisms: dislocation-mediated plastic flow and atomic surface diffusion. To elucidate the operative mechanism under different thermal conditions, we employed Mean Square Displacement (MSD) analysis to track atomic mobility and DXA to quantify plastic deformation. Our analysis reveals a distinct, temperature-dependent crossover from a plasticity-dominated to a diffusion-dominated regime.

\begin{figure}[!t]
    \centering
    \includegraphics[width=0.75\textwidth]{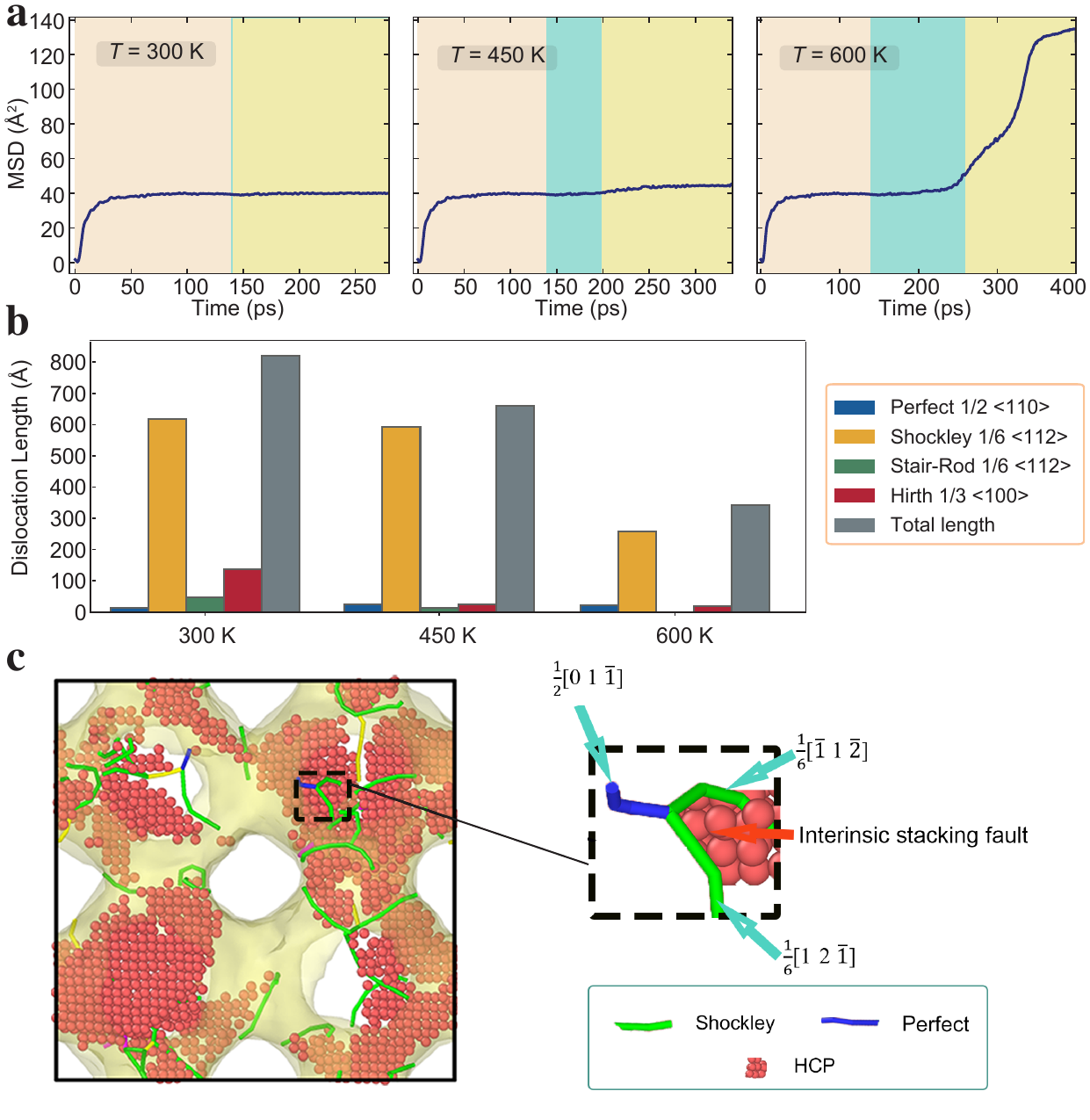}
    \caption{\textbf{Temperature-dependent sintering mechanism of \alcu NPs.} 
    (a) Mean square displacement (MSD) of all atoms during sintering at 300, 450, and 600 K, calculated with respect to the initial configuration. 
    (b) Dislocation length evaluated at the midpoint of the holding stage. 
    (c) DXA-revealed microstructure of the \alcu NPs at 450 K during the holding stage, highlighting the evolution of the dislocation network.}
    \label{fig7}
\end{figure}

At the highest sintering temperature of 600~K, we observe a dramatic increase in atomic mobility. The MSD curve exhibits a steep slope, culminating in a maximum atomic displacement of approximately 140 \AA$^2$ during the holding stage (\Fig{fig7}a). This value is more than three times greater than that observed at lower temperatures, providing clear evidence of a highly active, liquid-like diffusion process characteristic of a system approaching its surface melting point. Concurrently, DXA reveals a significant suppression of dislocation activity, with the total dislocation length reaching its minimum value across all tested conditions (\Fig{fig7}b). The elevated thermal energy actively promotes recovery mechanisms that annihilate dislocations, effectively shutting down the plastic flow pathway. Together, these results unambiguously establish that sintering at 600 K is governed by atomic surface diffusion.

In contrast, at 300 K and 450 K, atomic mobility is substantially curtailed, with maximum MSD values reaching only $\sim40-43$~\AA$^2$ (\Fig{fig7}a). This limited atomic displacement indicates that surface diffusion is kinetically hindered and cannot be the primary driver of consolidation. Instead, DXA analysis at these temperatures reveals a robust dislocation-mediated mechanism. We observe the nucleation and glide of Shockley partial dislocations, which subsequently combine to form perfect dislocations in the presence of stacking faults (\Fig{fig7}b,c). This process, a hallmark of plastic deformation, provides the necessary mass transport for neck growth in the absence of significant thermal diffusion. The high stacking fault energy of aluminum facilitates this dislocation activity, making it a highly efficient mechanism at lower temperatures~\cite{shih2021stacking,malti2025microstructural,linda2022effect}.

Taken togheter, our simulations have identified a clear mechanistic crossover controlled by thermal energy. At low-to-moderate temperatures ($\leq$ 450~K), where atomic mobility is limited, NP sintering proceeds via dislocation-mediated plastic flow. However, upon crossing a critical thermal threshold (approaching 600~K), a transition occurs: thermally activated recovery processes suppress dislocation activity, while a highly mobile, liquid-like surface layer emerges, shifting the dominant mechanism entirely to atomic diffusion.

\subsection{Kinetic Control of Final Nanostructure via Cooling Rate}
\begin{figure}[!t]
    \centering
    \includegraphics[width=\textwidth]{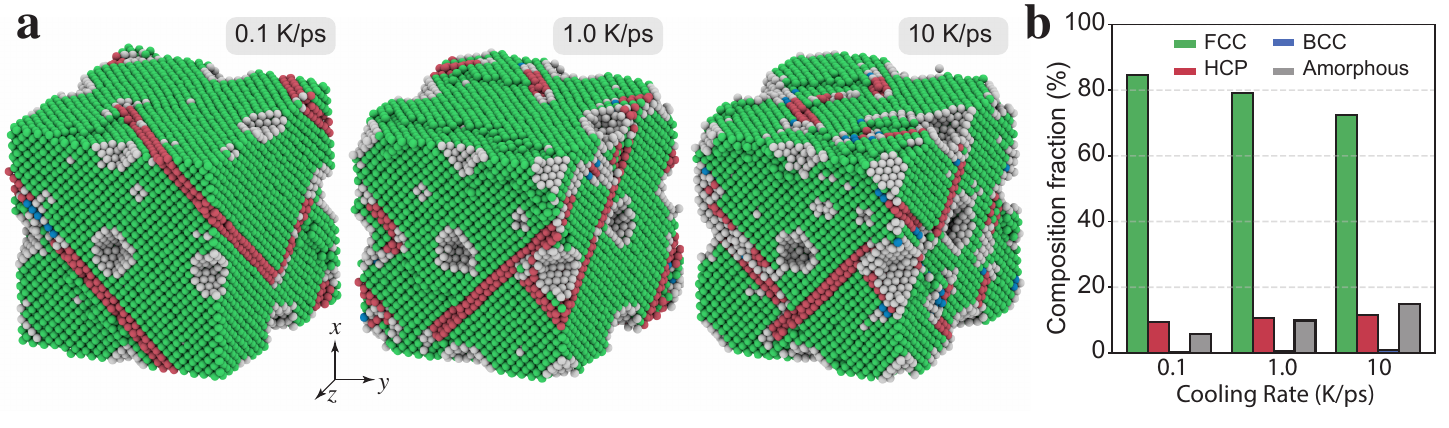}
    \caption{\textbf{Effect of cooling rate on microstructural evolution in Al–6.8\%Cu NPs.} 
    \textbf{a,} Final atomic configurations of the NPs after the cooling stage at rates of 0.1 K ps$^{-1}$, 1 K ps$^{-1}$, and 10 K ps$^{-1}$, respectively. 
    \textbf{b,} Quantitative common neighbor analysis (CNA) results for the corresponding final sintered Al–6.8\%Cu NPs shown in panel a. 
    }
    \label{fig9}
\end{figure}

The final atomic configuration of a sintered material is not determined solely by the peak temperature, but by the kinetic pathway taken to cool it. To isolate this effect, we performed three additional MD simulations in which the fully integrated nanostructure was subjected from the 600 K sintering process to three distinct cooling rates: a rapid quench (10~K/ps), a moderate cool (1~K/ps), and a near-equilibrium cool (0.1~K/ps), \Fig{fig9}b. Our results demonstrate that the cooling rate provides a powerful kinetic lever to precisely tune the balance between crystalline order, quenched-in disorder, and internal defect structures.
The primary competition during cooling is between thermodynamically-driven crystallization and kinetic trapping. As thermal energy is removed, the system seeks its lowest-energy state, the FCC lattice. At a slow cooling rate of 0.1 K/ps, the system is afforded sufficient time for atomic rearrangement, resulting in a highly crystalline structure with a minimal residual amorphous fraction of just 5.8\% (\Fig{fig9}b). However, as the cooling rate increases, atoms are increasingly unable to diffuse to their ideal lattice sites before the structure is frozen in place. This kinetic trapping results in a progressively higher fraction of the high-temperature amorphous phase being retained, reaching a maximum of 14.8\% under the rapid 10 K/ps quench.

Beyond the degree of crystallinity, the cooling rate also dictates the density of planar defects within the crystalline domains. These defects, which manifest as HCP stacking faults, serve as a secondary mechanism for energy reduction during the rapid solidification process~\cite{he2018situ}. While a baseline of 9.4\% HCP is observed at the slowest cooling rate, this fraction increases to 11.7\% at the fastest rate (\Fig{fig9}a,b). This trend indicates that when the system lacks the time for perfect FCC ordering, it increasingly relies on the formation of these lower-energy planar defects. This finding is significant, as it establishes that the internal defect landscape—a critical factor for mechanical properties—is not only a function of peak sintering temperature but is also directly controlled by the subsequent cooling kinetics.
Hence, the cooling process is not a passive step but an active instrument for nanostructural design. By controlling the cooling rate, one can precisely manipulate the final state of the material, dialing in a desired ratio of crystalline-to-amorphous content and engineering the internal defect density.

\section*{Conclusion}

In this work, we have demonstrated that the consolidation of metallic nanoparticles is not a simple process of diffusional bonding but a sophisticated, multi-stage transformation governed by a delicate interplay of thermodynamics and kinetics. The main discovery of this study is the existence of a mechanistic crossover dictated by the sintering temperature. At low-to-moderate temperatures ($\leq 450$~K), where atomic mobility is constrained, sintering proceeds primarily through dislocation-mediated plastic flow. In this regime, consolidation is a purely solid-state mechanical process. However, upon reaching a critical thermal threshold (approaching 600~K, or $\sim 0.8\,T_\mathrm{m}$), the system undergoes a fundamental change. A transient, liquid-like amorphous phase emerges, and thermally-activated recovery processes annihilate dislocations. This triggers a crossover to a new dominant mechanism: atomic surface diffusion, which facilitates rapid mass transport, void elimination, and the formation of a dense, fully integrated nanostructure.
Following consolidation, the cooling pathway provides a second, kinetic lever for tuning the final material properties. Our results show that the cooling rate enables precise manipulation of the final crystallographic state. Slow, near-equilibrium cooling promotes a transition to the lowest-energy, highly crystalline FCC state, whereas rapid quenching kinetically traps a significant fraction of the high-temperature disordered phase and increases the density of planar stacking-fault defects.

Ultimately, this research establishes a clear processing, mechanism, structure relationship. By understanding the distinct roles of peak temperature in selecting the dominant physical mechanism and of the cooling rate in governing the final atomic arrangement, we can move beyond empirical trial-and-error. This work provides a predictive roadmap for the rational design of advanced nanomaterials, enabling the engineering of specific microstructures—from highly crystalline to partially amorphous, with tailored defect landscapes—to achieve desired functional properties.

\section*{Acknowledgments}
\noindent I.A.A. appreciates the financial support from the Simulation Science Center Clausthal/Göttingen, Germany and the German Research Foundation (DFG) under contract GU1530/11-1, SPP 2315.

\clearpage
\newpage
\appendix

\setcounter{figure}{0}
\renewcommand{\thefigure}{A.\arabic{figure}}
\setcounter{table}{0}
\renewcommand{\thetable}{A.\arabic{table}}

\section{Supplementary Information}
\begin{table}[h!]
\centering
\caption{Sintering parameters of \alcu NPs used in this study.}
\begin{tabular}{cccc}
\hline
\textbf{Temperature (K)} & \textbf{Heating rate (K/ps)} & \textbf{Holding time (ps)} & \textbf{Cooling rate (K/ps)} \\
\hline
300 & N/A & 140 & N/A \\
450 & 2.5 & 140 & 1 \\
600 & 2.5 & 140 & 0.1, 1, 10 \\
\hline
\end{tabular}
\label{tab:sintering_params}
\end{table}

\begin{figure}[!h]
    \centering
    \includegraphics[width=0.7\textwidth]{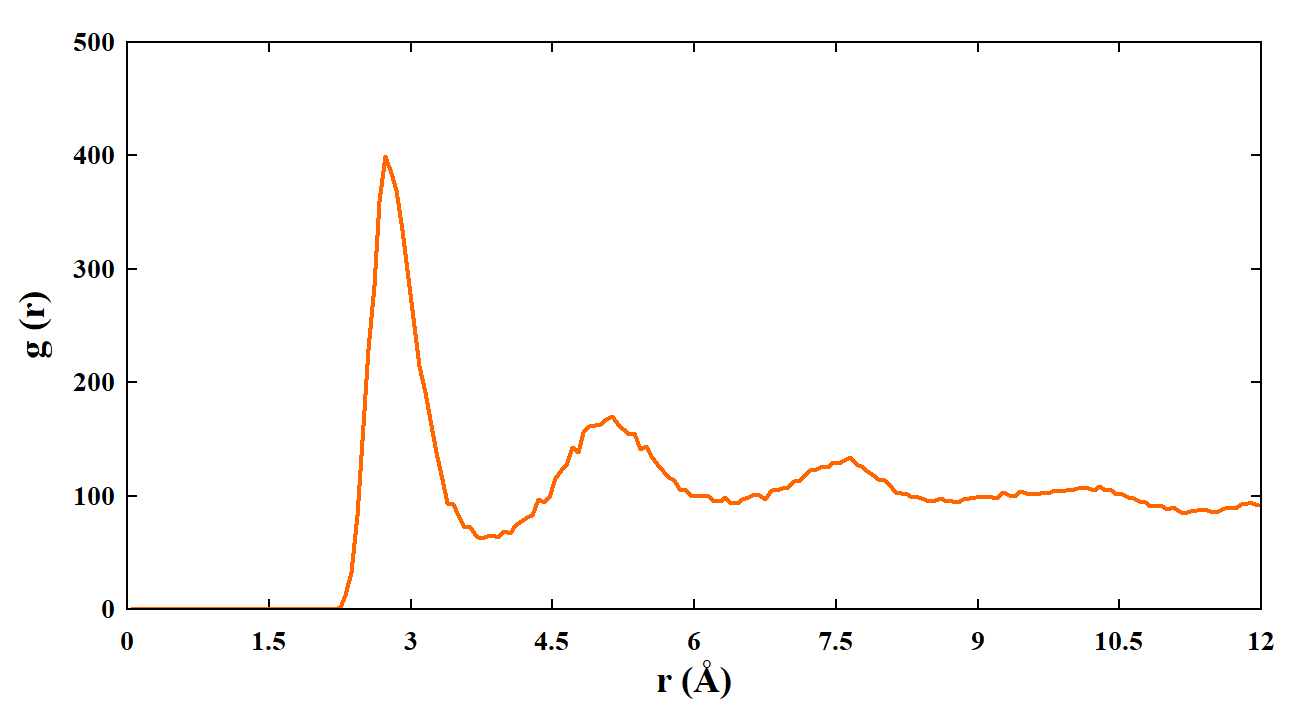}
    \caption{\textbf{Radial distribution function (RDF) of amorphous atoms in the \alcu NPs at 300~K.}
    The RDF, $g(r)$, characterizes the distribution of interatomic distances within the system, reflecting the local structural ordering of atoms.
    It was computed only for the amorphous atoms of the Al–6.8\%Cu NPs at 300~K to evaluate deviations from the ideal face-centered cubic (FCC) structure.
    The broadening and reduction of peak intensities indicate the loss of long-range order while retaining short-range coordination typical of metallic systems.}
    \label{SI_rdf}
\end{figure}

\begin{figure}[!h]
    \centering
    \includegraphics[width=0.5\textwidth]{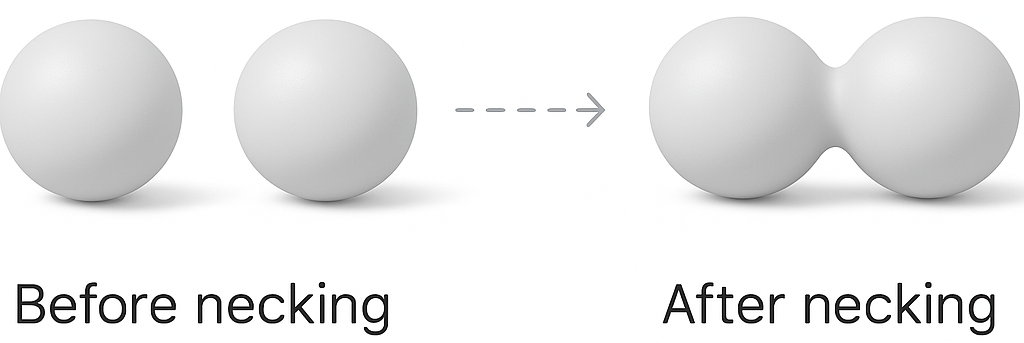}
    \caption{\textbf{Schematic illustration of NPs necking during sintering.} Two individual NPs (left) form a neck region as they coalesce under thermal activation (right), representing the initial stage of particle fusion.}
    \label{SI_necking}
\end{figure}

\end{document}